\documentclass[twocolumn,english,aps,prb,showpacs,byrevtex,amsmath,amssymb,superscriptaddress,longbibliography]{revtex4-1}

\usepackage{graphicx}
\usepackage{epstopdf}
\usepackage{dcolumn}
\usepackage{bm}
\usepackage{gensymb}
\usepackage{upgreek}
\usepackage{hyperref}
\usepackage{graphicx}
\usepackage{dcolumn}
\usepackage{bm}
\usepackage{gensymb}
\usepackage{upgreek}
\usepackage{booktabs}
\usepackage{hyperref}

\usepackage{amssymb,mathtools}
\usepackage{color}
\usepackage{amsmath} 

\usepackage{tikz,xcolor,hyperref}

\definecolor{lime}{HTML}{A6CE39}
\DeclareRobustCommand{\orcidicon}{%
	\begin{tikzpicture}
	\draw[lime, fill=lime] (0,0)
	circle [radius=0.16]
	node[white] {{\fontfamily{qag}\selectfont \tiny ID}};
	\draw[white, fill=white] (-0.0625,0.095)
	circle [radius=0.007];
	\end{tikzpicture}
	\hspace{-2mm}
}

\foreach \x in {A, ..., Z}{%
	\expandafter\xdef\csname orcid\x\endcsname{\noexpand\href{https://orcid.org/\csname orcidauthor\x\endcsname}{\noexpand\orcidicon}}
}

\begin{document}

\title{Staggered Dzyaloshinskii-Moriya inducing weak ferromagnetism in centrosymmetric altermagnets and weak ferrimagnetism in noncentrosymmetric altermagnets}

\author{Carmine Autieri\orcidA}
\email{autieri@magtop.ifpan.edu.pl}
\affiliation{International Research Centre Magtop, Institute of Physics, Polish Academy of Sciences, Aleja Lotnik\'ow 32/46, 02668 Warsaw, Poland}

\author{Raghottam M Sattigeri\orcidC}
\email{raghottam.sattigeri@unimi.it}
\affiliation{Physics Department, Universit\'a degli Studi di Milano, Via Celoria 16, 20133 Milan, Italy}

\author{Giuseppe Cuono\orcidD}
\email{gcuono@magtop.ifpan.edu.pl}
\affiliation{International Research Centre Magtop, Institute of Physics, Polish Academy of Sciences, Aleja Lotnik\'ow 32/46, 02668 Warsaw, Poland}

\author{Amar Fakhredine\orcidB}
\affiliation{Institute of Physics, Polish Academy of Sciences, Aleja Lotnik\'ow 32/46, 02668 Warsaw, Poland}

\begin{abstract}
The Dzyaloshinskii-Moriya interaction (DMI) has explained successfully the weak ferromagnetism in {\it some} centrosymmetric antiferromagnets. However, in the last years, it was generally claimed that the DMI is not effective in such systems. We reconciled these views by separating the conventional antiferromagnets from altermagnets. 
Altermagnets are collinear magnets having zero magnetization preserved by crystal symmetries in the non-relativistic limit. The spin-up and spin-down sublattices can be connected either by a proper rotation or by a combination of a rotation and a mirror or an inversion symmetry. Consequently, the system shows even-parity wave spin order in the k-space lifting the Kramer's degeneracy in the non-relativistic band structure leading to unconventional magnetism.
The DMI can create weak ferromagnetism or weak ferrimagnetism in centrosymmetric and in noncentrosymmetric altermagnets while it is not effective in conventional antiferromagnets.  
Once the spin-orbit coupling is included in an altermagnetic system (where the time-reversal symmetry is broken), the components of spin moments of the two sublattices along the N\'eel vector are antiparallel but the other two spin components orthogonal to the N\'eel vector can be null, parallel or antiparallel. In cases where we have different bands showing parallel and antiparallel spin components at the same time, the magnetic order results in weak ferrimagnetism. If we restrain to high-symmetry directions for the N\'eel vector, we find weak ferrimagnetism only in the noncentrosymmetric compound MnSe among the examined cases. The altermagnetic compounds can host weak ferromagnetism, weak ferrimagnetism or have zero magnetization. Restricted to the cases considered in this paper, the Hall vector is orthogonal to the N\'eel vector in the case of weak ferromagnetism while in the case of weak ferrimagnetism the Hall vector has a component parallel to the N\'eel vector too. While the weak ferromagnetism has net magnetization with components linearly proportional to the DMI, the weak ferrimagnetism in MnSe shows components proportional to the second-order.
We find a sign change of the magnetization, and possibly of the anomalous Hall effect, as a function of the band filling and N\'eel vector. We describe the dependence of the weak ferromagnetism on the charge doping. 
\end{abstract}

\pacs{}

\maketitle
 
\section{Introduction}
Abundant studies have emerged in the past years motivated by the phenomena of time-reversal symmetry breaking and spin splittings of the band structures\cite{doi:10.1126/sciadv.aaz8809,hayami2019momentum,hayami2020bottom,mazin2021prediction,yuan2021prediction}, which are originally properties of a ferromagnet. Nevertheless, it was recently discovered that non-relativistic spin-splitting in electronic bands can be a feature in antiferromagnetic materials having two equivalent sublattices with magnetic moments related by crystal-symmetry\cite{Smejkal22beyond}. This newly identified magnetic phase has been named altermagnetism. (AM)\cite{Smejkal22,mazin2022altermagnetism,Autieri2024}
The existence of AM necessitates that the electronic charge of spin-up (down) atoms should be mapped into spin-down (up) atoms not through translation or inversion as in the case of conventional antiferromagnets, but through the application of rotations, mirrors, or combination of a rotation and an inversion symmetry. In another study, the altermagnetic materials have been dubbed antiferromagnetism with non-interconvertible spin-structure motif pair\cite{yuan2023degeneracy}. As a result, AM can be observed in specific space groups, and more precisely, in the magnetic space groups of type-I and type-III\cite{GUO2023100991}.
Numerous transition metal oxides\cite{GUO2023100991} and rare-earth compounds have been documented to present altermagnetism\cite{Cuono23EuCd2As2}. 
Altermagnets have a wide range of potential applications, they are promising to enable highly efficient spin-current generation\cite{Hernandez21}, as well as contributing to the development of giant and tunneling magnetoresistance\cite{GiantMagneto22}. It was shown that spin-independent conductance in altermagnets can be efficiently exploited in spintronics \cite{Shao21,shao2023neel}. Furthermore, altermagnets can play a significant role in spincaloritronics\cite{zhou2023crystal} and can be of practical use in Josephson junctions\cite{Ouassou23}.

\begin{figure}[ht!]
\centering
\includegraphics[width=\linewidth]{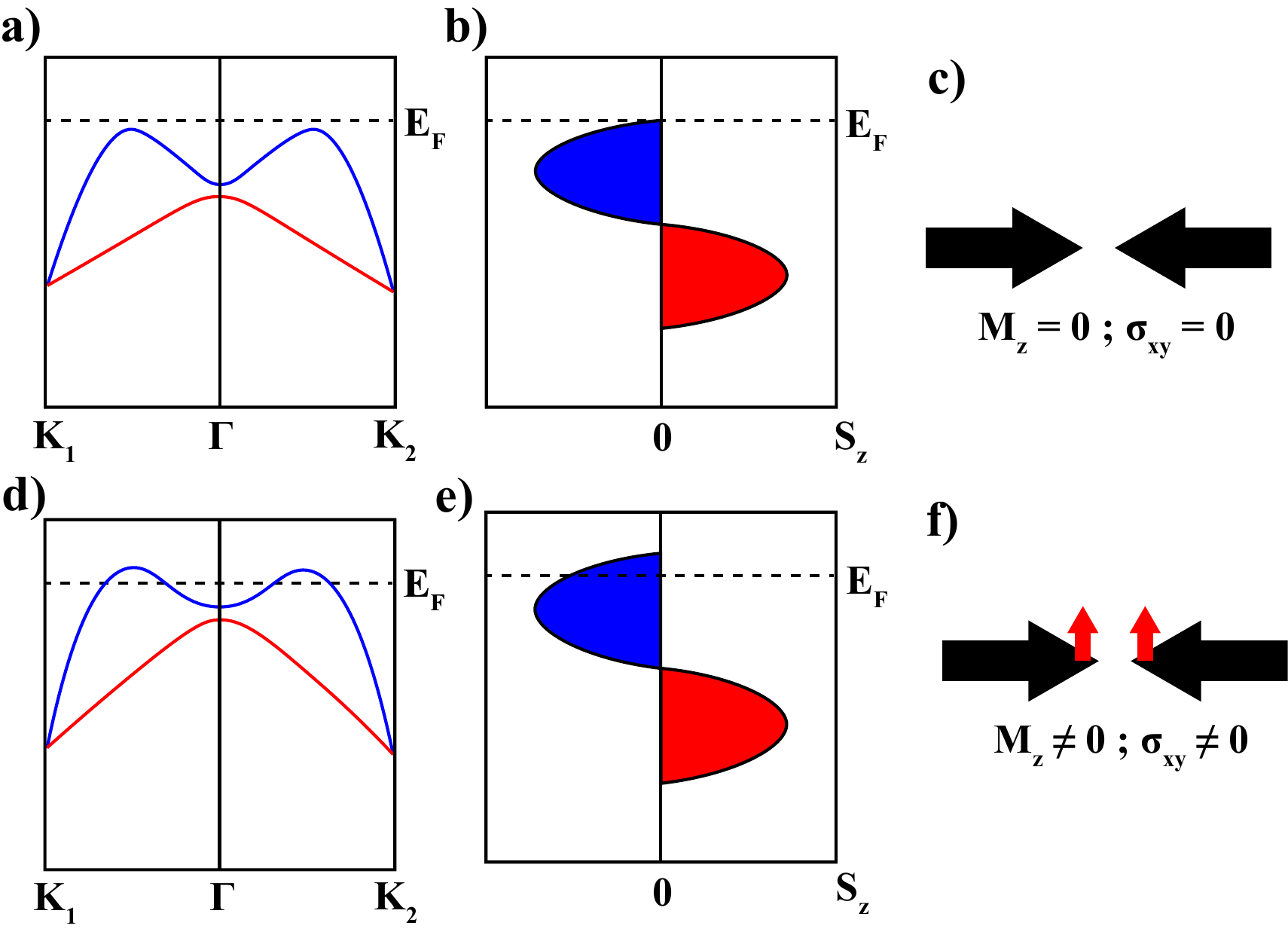}
\caption{Origin of weak ferromagnetism in altermagnets. (a) Schematic band structure of an altermagnet with SOC, N\'eel vector in the xy plane and Fermi level in the gap. The red and blue colors represent the negative and positive S$_z$ component of the total system. (b) Local S$_z$ component for the magnetic atom as a function of the energy for the energy spectrum (a), in the centrosymmetric case  S$_z$ for the two atoms are equal. (c) Magnetic configuration without M$_z$ and AHE $\rho_{xy}$. (d) Same as (a) with metallicity. (e) S$_z$ component as a function of the energy for the energy spectrum (d). (f) Magnetic configuration with non-zero M$_z$ components and non-zero AHE $\rho_{xy}$.}\label{Fig1}
\end{figure}

The spin-orbit coupling preserves the time-reversal symmetry, the product of time-reversal symmetry with inversion or translation connecting sublattices with opposite spin. Therefore, in a system with Kramer’s degeneracy protected by those symmetries, the spin-orbit cannot create any net magnetization or weak ferromagnetism.
In the case of altermagnetic compounds, we have a breaking of the time-reversal symmetry and the spin-orbit could generate the so-called weak ferromagnetism\cite{DZYALOSHINSKY1958241}. 
The presence of the altermagnetic spin-splitting is therefore a necessary condition to obtain weak ferromagnetism.
For historical reasons, the rise of weak ferromagnetism is realized by introducing the Dzyaloshinskii-Moriya interaction (DMI).
The DMI is a relativistic antisymmetric component of exchange interaction that occurs due to spin-orbit coupling.  It is responsible for spin-canting which may result in the formation of magnetic chiral structures\cite{guo2017spin}, and therefore it was used to explain "weak ferromagnetism" in materials like the centrosymmetric antiferromagnet Fe$_2$O$_3$\cite{camley2023consequences,DZYALOSHINSKY1958241,PhysRev.120.91} with space group 167. This weak ferromagnetism has a relativistic origin and is the only kind of weak ferromagnetism discussed in this paper. 
We can also have the rotation of the N\'eel vector when the spin-orbit induces antiferromagnetic components of the spin\cite{PhysRevB.62.14229}.
Other centrosymmetric compounds belonging to the space group 62 exhibit weak ferromagnetism and lead to a significant ferromagnetic behavior, which was later explained by the mechanism of the Dzyaloshinskii Moriya exchange based on torque measurements\cite{salazar2022presence,treves1962magnetic}. 
The spin canting has been attributed to the DMI in most cases, although other effects share the same requirements to be imposed on the crystal symmetry\cite{zhou2020weak}. The weak ferromagnetism can also be present in complex systems like solids with molecules as cations\cite{Blake79} and biphasic systems\cite{Koval17}.
Recently, the presence of weak ferromagnetism and weak ferrimagnetism was highlighted also in 2D altermagnets\cite{milivojevic2024interplay}. 
In other literature, the weak ferromagnetism from DMI in bulk systems was associated with noncentrosymmetric systems\cite{Weng11}. Furthermore, in ferroelectric BiFeO$_3$, a subtle DMI which constitutes about 1\% of the antiferromagnetic exchange strength causes a tilting of the two spin sublattices, consequently giving rise to "weak ferromagnetism"\cite{camley2023consequences,dixit2015stabilization}. 
The Dzyaloshinskii-Moriya vector between the sites $i$ and $j$ is $\vec{D}_{i,j}$ while the DMI has the antisymmetric form of $\vec{D}_{i,j}\cdot(\vec{S}_{i}\times\vec{S}_{j})$
where S$_i$ and S$_j$ are the spins on sites i and j. 
In case of broken inversion symmetry, the total DMI $\sum_{i,j}\vec{D}_{i,j} \ne$0\cite{PhysRevLett.115.267210}. 
In the case of crystal structures with inversion symmetry we have that $\sum_{i,j}\vec{D}_{i,j}$=0, if the system has also inversion symmetry in the mean point between the sites i and j as in conventional antiferromagnets, the DM vectors are all zero $\vec{D}_{i,j}$=0. However, if the sites i and j are connected by rototranslation as in altermagnets, the DM vectors are not zero $\vec{D}_{i,j}\ne$0.
Hence, altermagnets with inversion symmetry have $\sum_{i,j}\vec{D}_{i,j}$=0 and $\vec{D}_{i,j}\ne$0, which results in a staggered DMI. Therefore, differently from what is usually claimed in the literature\cite{CAMLEY2023100605}, the crystal structures supporting DMI do not necessarily display broken inversion symmetry\cite{weng2011framework}.
The total DMI plays an important role in forming skyrmions with a fixed chirality and vorticity in the case of noncentrosymmetric magnets since it allows the tilting of adjacent spins under special rules related to crystal symmetry. In contrast, in centrosymmetric structures, other mechanisms are believed to be responsible for the complex magnetism\cite{tokura2020magnetic}.

\begin{figure*}[ht!]
\centering
\includegraphics[width=\linewidth]{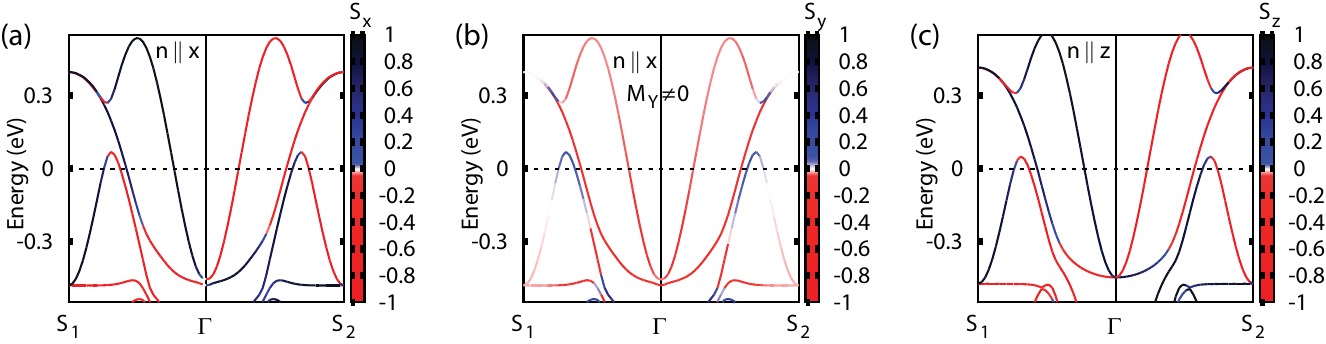}
\caption{Band structure of d-wave RuO$_2$ along the S$_1$-$\Gamma$-S$_2$ low-symmetry k-path. (a) The positive and negative S$_x$ components are plotted in red and blue, respectively, while the N\'eel vector is along the x-axis. (b)  The positive and negative S$_y$ components are plotted in red and blue, respectively, while the N\'eel vector is along the x-axis. (c) The positive and negative S$_z$ components are plotted in red and blue, respectively, while the N\'eel vector is along the z-axis. S$_x$, S$_y$ and S$_z$ represent the spin components for the total system.
The Fermi level is set to zero. The coordinates of the k-points are S$_1$=(0.5,0.5,0) and S$_2$=(-0.5,0.5,0) in direct coordinates.}
\label{RuO2_S}
\end{figure*}

The non-relativistic spin-splitting is proportional to the hopping parameters which tend to increase with the atomic number. Large spin-splittings have been found in compounds with heavy elements such as MnTe, RuO$_2$, CrSb, etc ... which are also the most representative and most studied compounds among the altermagnetic systems. 
Beyond the large spin-splitting, these heavy atoms own a huge spin-orbit coupling (SOC). Therefore, the study of the spin-orbital effects is extremely relevant, especially for the most representative altermagnets. Beyond that, the anomalous Hall effect (AHE) strongly depends on the size of the SOC.
In the case of a metallic system, the DMI can produce weak ferromagnetism in AM compounds that display spontaneous AHE\cite{doi:10.1126/sciadv.aaz8809,PhysRevLett.130.036702} which can be enhanced by Weyl points\cite{10.1063/5.0158271}.
The orientation of the Hall vector strongly depends on the N\'eel vector, which is defined as the difference between the magnetization vectors of the two distinct magnetic sublattices\cite{turek2022altermagnetism,shao2023neel, Fakhredine23}. 
It was predicted that the AHE in altermagnets is non-linear, where the Hall voltage is not linear to the applied electric field. In such systems, the AHE has been proven to be able to detect the N\'eel vector. \cite{PhysRevLett.124.067203,fang2023quantum,vsmejkal2022anomalous,godinho2018electrically}
While the direction of the Hall vector has been explored using symmetry considerations, there is a notable absence of information in the literature regarding how to establish this orientation based on fundamental electronic properties through density functional theory calculations. 
The topological aspects were investigated in altermagnets and they are more likely to appear in the absence of weak ferromagnetism. GdAlSi is a Weyl altermagnet\cite{nag2023gdalsiantiferromagnetictopologicalweyl}, RuO$_{2}$ shows a Kramer nodal-line and XMene can host altermagnetic quantum spin insulator\cite{antonenko2024mirrorchernbandsweyl}. Although these topological phases are all different, interestingly all cases required the N\'eel vector along the z-axis to remove the weak ferromagnetism and to keep a high symmetry which seems necessary for topological protection.

In this work, our objective is to have a general understanding of the appearance of weak ferromagnetism and weak ferrimagnetism in centrosymmetric and noncentrosymmetric systems.
We discuss the appearance of weak ferromagnetism and weak ferrimagnetism as relativistic effects which could also be properties of altermagnets. Once the time-reversal symmetry is broken and the Kramers degeneracy is lifted, the spin components of the two sublattices orthogonal to the N\'eel vector are related by symmetry in centrosymmetric systems and therefore must be parallel or antiparallel, if they are not null. If the orthogonal components end up antiparallel, this would correspond to an overall rotation of the N\'eel vector. The scenario changes in the noncentrosymmetric system MnSe where different sets of bands can generate at the same time parallel or antiparallel spin components in which we establish that the direction of the Hall vector is orthogonal to the N\'eel vector.
A schematic representation of our main results is presented in Fig. \ref{Fig1}. We propose that the band structure reported in Fig. \ref{Fig1}(a) can be present in all altermagnets for a suitable N\'eel vector, as a consequence of this band structure, the local density of states (DOS) for the component of the spin orthogonal to the N\'eel vector have a relativistic spin-splitting as shown in \ref{Fig1}(b). With the Fermi level in the band gap with the filled electronic shell, we still observe a net zero magnetic configuration M$_z$ and zero anomalous Hall effect in altermagnets. Even if it is allowed by symmetry, the weak ferromagnetism is absent in systems with filled electronic shells, while the weak ferromagnetism is absent or strongly suppressed when the electronic occupation is at half-filling probably due to the zero angular momentum on which the DMI depends\cite{PhysRev.120.91}. The weak ferromagnetism and the anomalous Hall effect are both axial vectors and therefore allowed by the same magnetic symmetries. However, when bands cross the Fermi level (as in Fig. \ref{Fig1}(d)) which is also evident from the increase in the contribution from the S$_z$ component (as in Fig. \ref{Fig1}(e)), the SOC creates an effective non-zero magnetic configuration giving rise to ferromagnetic spins aligned perpendicular to the N\'eel vector (as in Fig. \ref{Fig1}(f)). By integrating in energy the spin component orthogonal to the N\'eel vector in a realistic multiband system, we will obtain the magnetization as a function of the energy, this magnetization can be zero at the same energies by compensation of the spins, however, the compensation of the spins does not mean that we have a compensation of the Berry curvature. Therefore, a null weak magnetization is not associated with a null anomalous Hall effect\cite{PhysRevResearch.2.043394}. Here, it is worth mentioning that the AHE is not a direct consequence of a weak ferromagnetic behavior even though they share the same requirements to exist\cite{PhysRevLett.130.036702}. The weak ferromagnetism is difficult to detect especially on the nanoscale level, so our work provides a guide to theoretically predict the direction of the weak ferromagnetism to support the experimental investigation.
Using density functional theory calculations, the results of the self-consistent calculation will report parallel or antiparallel components depending on which the total energy is minimized. We calculate the spin components of the two sublattices as a function of the N\'eel vector for the centrosymmetric d-wave RuO$_2$, the centrosymmetric g-wave CrSb and the noncentrosymmetric g-wave MnSe. 
The paper is organized as follows: the second Section is dedicated to the results of the three chosen compounds. The third Section shows the relevance of the staggered DMI to the rise of weak ferromagnetism. Finally, in the fourth Section, the authors draw their conclusions.

\section{Results}

In a centrosymmetric conventional antiferromagnet, we have Kramers degeneracy therefore the DOS of the spin-up sublattice and the spin-down sublattice are always equal for all components. The two sublattices have always antiparallel spin components and no weak ferromagnetism is possible.  In a noncentrosymmetric conventional antiferromagnet with SOC, it is possible to have an additional spin-splitting. this spin splitting can be Rashba-like or have other characteristics but the spin components belonging to two sublattices are antiparallel.\cite{PhysRevB.100.245115} Therefore, in noncentrosymmetric conventional antiferromagnets, there is no weak ferromagnetism as well.

We move to the study of the electronic properties responsible for weak ferromagnetism in centrosymmetric and noncentrosymmetric altermagnets.
We study the electronic properties of d-wave centrosymmetric altermagnet RuO$_2$ in the first subsection. In the second and third subsections, we compare the g-wave altermagnets CrSb and wurtzite MnSe, and we discuss the different effects of SOC.
All selected compounds have two atoms per unit cell which we will call M$_1$ and M$_2$ (M= Ru, Cr, Mn) to represent the properties of the two sublattices with opposite spins. 

The Hubbard U effects for the 4d-orbitals of Ru$^{4+}$, the 3d-orbitals of Cr and 3d-orbitals of Mn$^{2+}$ have been adopted\cite{Liechtenstein95density}. We have used the value of U = 2 eV \cite{Ivanov2016,PhysRevResearch.4.023256} keeping J$_H$ = 0.15U. Using different values of U will not affect the results of this paper, since our conclusions are based on symmetry aspects. However, when U is set to zero, we obtain that RuO$_2$ is not magnetic. More details can be found in the Appendix A.

\begin{figure}[ht!]
\centering
\includegraphics[height=1.2\linewidth]{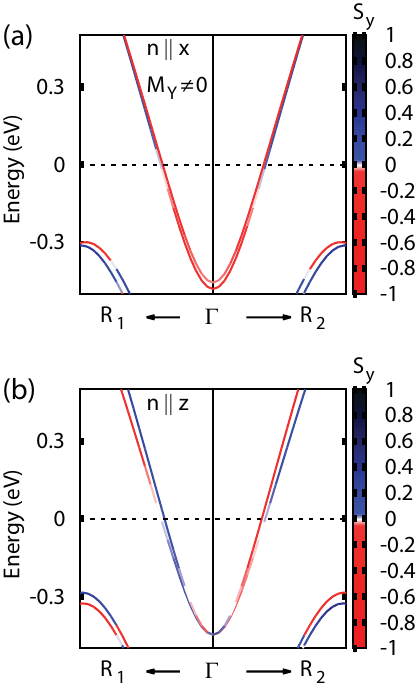}
\caption{Band structure of d-wave RuO$_2$ along the R$_1$-$\Gamma$-R$_2$ low-symmetry k-path with the N\'eel vector along (a) the x- and (b) z-axis. The positive and negative S$_y$ and S$_x$ components are plotted in red and blue, respectively. S$_x$, S$_y$ and S$_z$ represent the spin components for the total system. The Fermi level is set to zero. The coordinates of the k-points are R$_1$=(0.5,0,0.5) and R$_2$=(-0.5,0,0.5) in direct coordinates.}
\label{RuO2_R}
\end{figure}

\begin{figure*}[ht!]
\centering
\includegraphics[width=\linewidth]{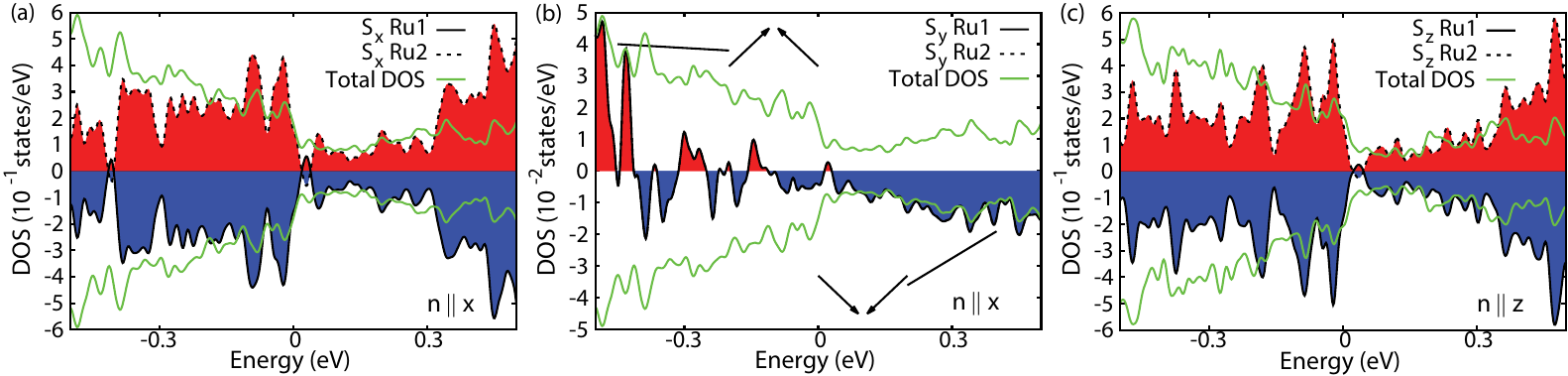}
\caption{DOS for the spin-components as a function of the energy for the d-wave RuO$_2$. (a,b) DOS of the S$_x$ and S$_y$ components with the N\'eel vector along the x-axis, respectively. For the DOS of the S$_y$ components, the positive (negative) DOS contributes to a weak ferromagnetism with positive (negative) S$_y$ as shown by the arrows representing the spins in the figure. (c) DOS of the S$_z$ component with N\'eel vector along the z-axis. The other components are antiparallel. The Fermi level is set to zero. The total DOS is plotted in green and renormalized by a factor of $10^{-1}$ states/eV to fit within the plot. The negative values of the total DOS correspond to spin down while the positive values correspond to spin up.}
\label{RuO2_DOS}
\end{figure*}

\begin{figure*}[ht!]
\includegraphics[width=\linewidth]{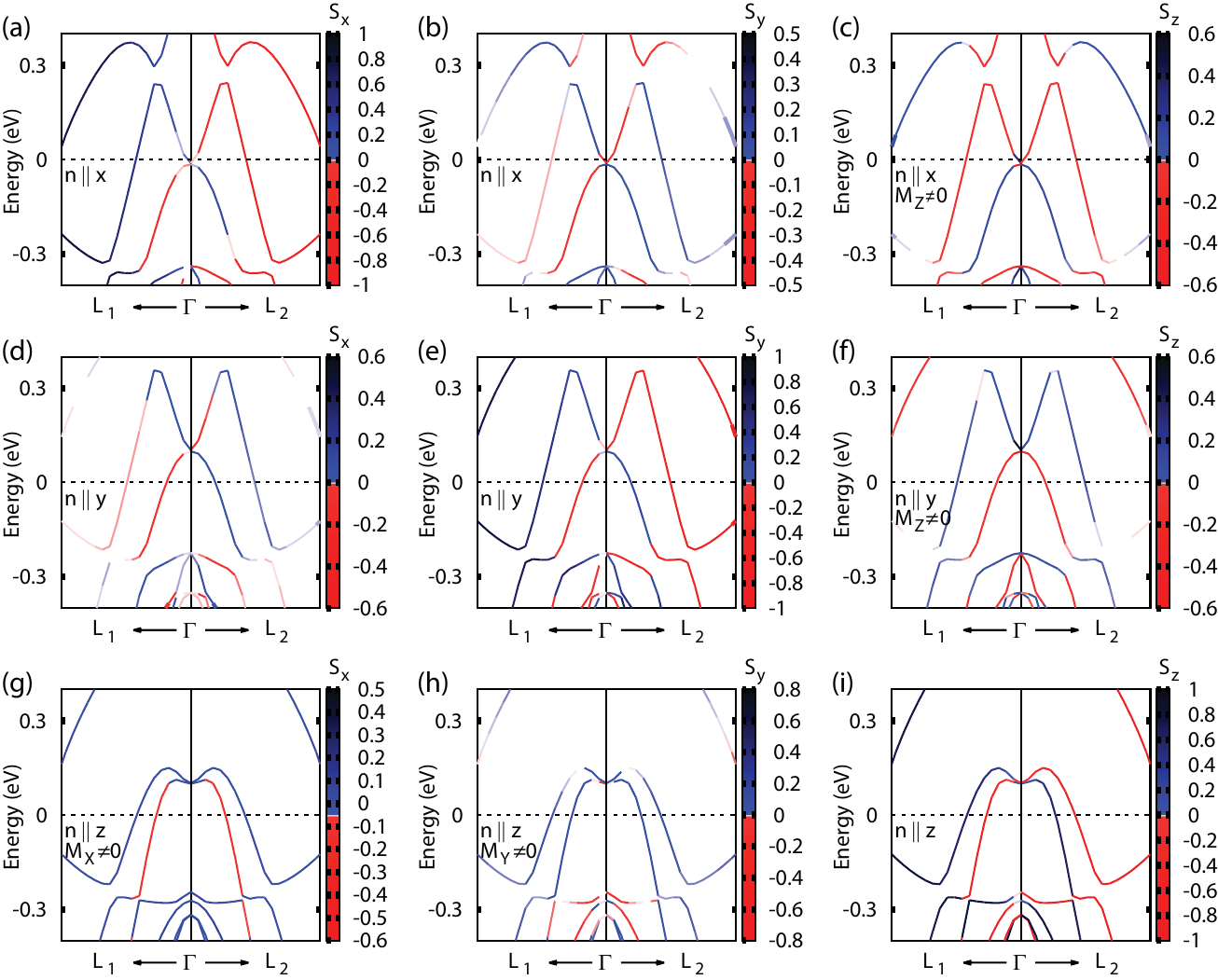}
\caption{Band structure of centrosymmetric CrSb with SOC along the L$_1$-$\Gamma$-L$_2$ low-symmetry k-path. With the N\'eel vector along the x-axis, we report the projection of the spin-component (a) S$_x$, (b) S$_y$ and (c) S$_z$. With the N\'eel vector along the y-axis, we report the projection of the spin-component (d) S$_x$, (e) S$_y$ and (f) S$_z$. With the N\'eel vector along the z-axis, we report the projection of the spin-component (g) S$_x$, (h) S$_y$ and (i) S$_z$. S$_x$, S$_y$ and S$_z$ represent the spin components for the total system. The Fermi level is set to zero. The coordinates of the k-points are L$_1$=(0.5,0,0.5) and L$_2$=(-0.5,0,0.5) in direct coordinates. When the \textbf{n} is along x, the total magnetization is M=(0,0,M$_z$). When the \textbf{n} is along y, the total magnetization is M=(0,0,M$_z$). When the \textbf{n} is along z, the total magnetization is M=(M$_X$,M$_Y$,0).}\label{L_CrSb}
\end{figure*}

\begin{figure}[ht!]
\centering
\includegraphics[width=\linewidth]{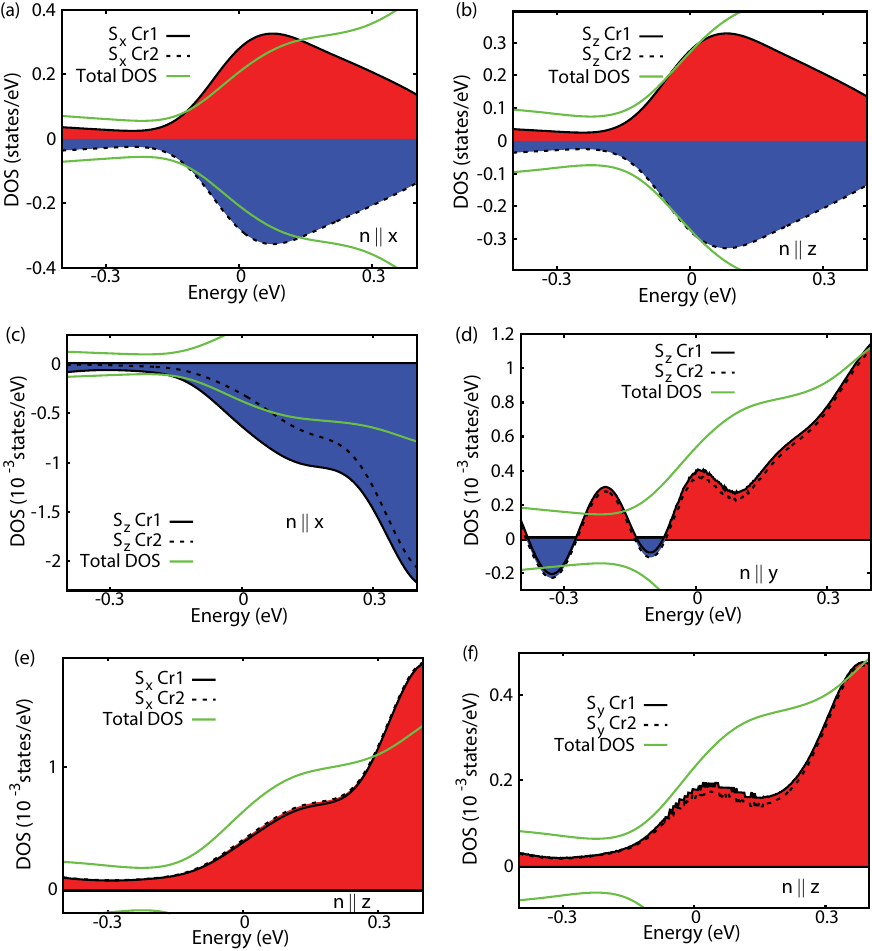}
\caption{DOS for the spin-components S$_x$, S$_y$ and  S$_z$ as a function of the energy for the centrosymmetric CrSb. (a,b) DOS for the components S$_x$ and  S$_z$ with the N\'eel vector along the x-axis and along the z-axis, respectively. (c,d) DOS for the spin component S$_z$ with the N\'eel vector along the x-axis and along the y-axis, respectively. Both of them show weak ferromagnetism. (e,f) DOS for the spin component S$_x$ and S$_y$ with the N\'eel vector along the z-axis. Both of them show weak ferromagnetism. The Fermi level is set to zero. The total DOS is renormalized by a factor of $10^{-3}$ states/eV} to fit within the plot. The negative values of the total DOS correspond to spin down while the positive values correspond to spin up.\label{DOS_CrSb}
\end{figure}

\subsection{Weak ferromagnetism in centrosymmetric tetragonal altermagnet RuO$_2$}

In this subsection, we describe the metallic centrosymmetric d-wave RuO$_2$. The altermagnetic properties of RuO$_2$ were investigated before \cite{doi:10.1126/sciadv.aaz8809,Feng2022,bai2022observation,karube2022observation,fedchenko2024observation}, however, our focus is on the band structure and the DOS for the spin-components. Since RuO$_2$ has open d-shells, we expect that this weak magnetization could be relatively large and could be calculated within DFT. Indeed, the magnetic moment orthogonal to the N\'eel vector is 0.022 $\mu_B$ per Ru atom along the y-axis when the N\'eel vector is along the x-axis.
In Figs. \ref{RuO2_S} and \ref{RuO2_R}, we report the band structures of RuO$_2$ projected on different spin components for two inequivalent directions of the N\'eel vector (x-axis and z-axis) along the S$_1$-$\Gamma$-S$_2$ and R$_1$-$\Gamma$-R$_2$ low-symmetry k-paths.
When the N\'eel vector is along the $x$-axis, the S$_x$ spin component of the total system for every energy has equal but opposite values along S$_1$-$\Gamma$ and $\Gamma$-S$_2$ as reported in Fig. \ref{RuO2_S}(a). The spin-splitting, in this case, is similar to the case of a non-relativistic spin-splitting. As a result, the magnetization component M$_x$ of the entire system that is the integral over the energy up to the Fermi level of S$_x$ is always zero. On the contrary, the spin components S$_y$ of the total system show a band structure where the value of S$_y$ along  S$_1$-$\Gamma$ equals the value along $\Gamma$-S$_2$.
For every value of the energy, the value of the magnetization M$_y$ will be in general non-zero resulting in a weak magnetization in the $y$-direction as we can see from Figs. \ref{RuO2_S}(b) and \ref{RuO2_R}(a). The origin of the canted spin moment derives exclusively from the spin-orbit coupling but this picture is equivalent to the DMI picture\cite{moriya1960anisotropic}.\\

In the case of RuO$_2$ with spins along the z-axis, we have an antiparallel order of the spin components, as shown in Figs. \ref{RuO2_S}(c) and \ref{RuO2_R}(b). This is a case of antiferromagnet with rotation of the N\'eel vector\cite{Kipp2021}.
In Fig. \ref{RuO2_R}(b), we can observe a small spin-splitting of the S$_y$ component since S$_y$ is altermagnetic too. It was recently demonstrated that in the noncollinear compound MnTe$_2$, all spin components can generate three different momentum-dependent spin splitting.\cite{zhu2024observation} In the rest of the paper, we will focus only on the case where SOC would generate weak ferromagnetism.
The other results were not shown because they are either equivalent to the shown results or with null spin components. 
The band degeneracy is broken at $\Gamma$ for the N\'eel vector along the x-axis, but not for the N\'eel vector along the z-axis. Therefore, we have 
lifting of the degeneracy at $\Gamma$ when the weak ferromagnetism rises.

To further extend our consideration to the entire Brillouin zone, we also show the density of states for the non-zero and inequivalent spin components in Fig. \ref{RuO2_DOS} for the N\'eel vector along the x- and z-axis. 
When the N\'eel vector is along the x-axis, the local DOSs of the S$_x$ components for the two sublattices are equal and opposite as described by the local DOS related to atoms Ru$_1$ and Ru$_2$ in Fig. \ref{RuO2_DOS}(a), therefore, the S$_x$ components are equal and opposite for the two sublattices. 
From the local DOS of the S$_y$ component reported in \ref{RuO2_DOS}(b), it is further clear that there are equal magnetic moments along the y-axis on the Ru$_1$ and Ru$_2$ atoms resulting in a weak ferromagnetism. The DOSs of the spin component S$_y$ for the atoms Ru$_1$ and Ru$_2$ are perfectly equal, therefore, the size of the local magnetic moments of Ru$_1$ and R$_2$ are equal. Comparing the DOS of the spin-component S$_x$ with the total DOS in green, we can see a reduction of S$_x$ around -0.4 eV in Fig. \ref{RuO2_DOS}(a) together with the maximum of S$_y$ in Fig. \ref{RuO2_DOS}(b), therefore, there is less  S$_x$ and more S$_y$ due to a strong rotation (canting) of both spins from the x-axis to the y-axis at -0.4 eV.

In the case of the N\'eel vector along the z-axis, the DOS of the spin-component S$_Z$ relative to Ru$_1$ and Ru$_2$ are equal and opposite as it is shown in Fig. \ref{RuO2_DOS}(c), while the DOSs for S$_X$ and S$_Y$ are negligible. Therefore, the Ru spins perfectly compensate each other and have no S$_X$ and S$_Y$, consequently, there is no weak ferromagnetism in agreement with previous results\cite{Feng2022}. From the symmetry point of view, this is a direct consequence of the magnetic group symmetries for the N\'eel vector along the z-axis.
Therefore, from the DOS we can also understand that the system presents weak ferromagnetic features depending on the orientation of the N\'eel vector.

In centrosymmetric systems, it was proven that RuO$_2$\cite{Feng2022} displays weak ferromagnetism if the N\'eel vector is along the principal axis while it can also display weak ferrimagnetism if the N\'eel vector is along a generic direction. For the 2D Lieb lattice\cite{antonenko2024mirrorchernbandsweyl}, weak ferromagnetism is observed for N\'eel vector along the (110) direction while weak ferrimagnetism is observed along the (100) direction. In hexagonal systems with N\'eel vector along the z-axis, we will prove the presence of weak ferromagnetism when the system is centrosymmetric such as CrSb, while we will detect weak ferrimagnetism when the system is noncentrosymmetric as in the case of MnSe.

\subsection{Weak ferromagnetism in centrosymmetric hexagonal altermagnet CrSb}

We define the x-axis of CrSb as parallel to the first neighbor Cr-Cr distance and the y-axis as orthogonal to it in the ab plane.
In Figs. \ref{L_CrSb}, we report the band structures for the g-wave CrSb centrosymmetric altermagnet projected on different spin components (S$_x$, S$_y$ and S$_z$) for the three different directions of the N\'eel vector along the L$_1$-$\Gamma$-L$_2$ focusing on the region near $\Gamma$ point. The ground state is the altermagnetic phase with the N\'eel vector along the z-axis, we have an energy difference of 0.2 meV per formula unit with respect to the in-plane N\'eel vector.  
When the N\'eel vector is along the x-axis as in Fig. \ref{L_CrSb}(a,b,c) or along the y-axis as in Figs. \ref{L_CrSb}(d,e,f), we have weak ferromagnetism only for S$_z$ resulting in a non-zero value of M$_z$. When the N\'eel vector is along the z-axis, we have weak ferromagnetism for the S$_x$ and S$_y$ components resulting in a non-zero M$_x$ and M$_y$ as shown in \ref{L_CrSb}(g,h,i). Overall, we find a weak ferromagnetism out-of-plane when the N\'eel vector is in-plane and a weak ferromagnetism in-plane when the N\'eel vector is out-of-plane. 
Additionally, we find a strong change in the relativistic band structure as a function of the N\'eel vector, especially between in-plane and out-of-plane directions of the N\'eel vector. 
Qualitatively similar results are observed for the high-symmetry k-path K$_1$-$\Gamma$-K$_2$ (not reported) with the difference that the non-relativistic spin splitting is absent.
We check the electronic properties in the DOS as well where we have information along the entire Brillouin zone and not for a limited number of k-points.
The DOS for S$_x$ and S$_z$ when the N\'eel vector is along the x- and z-axis are reported in  \ref{DOS_CrSb}(a) and (b), respectively.
We compare the DOS of the spin components with the total DOS illustrated in green.

At energy above 0.2 eV in Fig. \ref{DOS_CrSb}(a) and (b), the spin-component resolved DOS with spins along the N\'eel vectors has less weight with respect to the total DOS (green line). This means that above 0.2 eV there is DOS from non-magnetic atoms or there is a rotation of the spins with the weight of the DOS moving towards the spin components orthogonal to the N\'eel vector, in other words, the spins are rotated (canted) by spin-orbit to create the weak ferromagnetism. 
When the N\'eel vector is along the x- and y-axis, we have non-zero magnetization along the z-axis. The values of DOS for S$_z$ are reported in \ref{DOS_CrSb}(c) and (d) for the N\'eel vector along the x- and y-axis, since they are different we expect an anisotropy of the AHE\cite{leiviska2024anisotropy}. The case with N\'eel vector along the x-axis gives the largest weak ferromagnetism for CrSb with a value of -0.002 $\mu_B$ per formula unit. 
Above 0.2 eV, we can see from the DOS that S$_z$ is negative (Fig. \ref{DOS_CrSb}(c)) or positive (\ref{DOS_CrSb}(d)) depending on whether the N\'eel is along the x- or y-axis. This means that by rotating the N\'eel vector in the xy plane, we will have a sign change of M$_z$. This sign change of M$_z$ could also produce a sign change of the anomalous Hall effect. Recent studies on the sign change of the AHE in ferromagnets have explained this effect which is relatively rare in experimental conditions\cite{PhysRevLett.127.127202,PhysRevB.107.085102}. 
Recently, the AHE effect was measured for Mn$_5$Si$_3$, however, despite the presence of anisotropy of the AHE, the sign of the AHE was the same for N\'eel vectors along the x- and y-axis for this compound \cite{leiviska2024anisotropy}.
When the N\'eel vector is along the z-axis of CrSb, we observe weak ferromagnetism for both spin components for S$_x$ and S$_y$ as reported in \ref{DOS_CrSb}(e) and \ref{DOS_CrSb}(f).

The same properties investigated for the metallic CrSb are valid for the insulating MnTe which becomes metallic after intrinsic p-doping\cite{kluczyk2023coexistence}. 
MnTe with in-plane N\'eel vector presents weak ferromagnetism along the z-axis while the (001) surface of MnTe presents uncompensated magnetization along the N\'eel vector due to uncompensated spins on the surface\cite{D3NR03681B}.

\begin{figure*}[ht!]
\includegraphics[width=\linewidth]{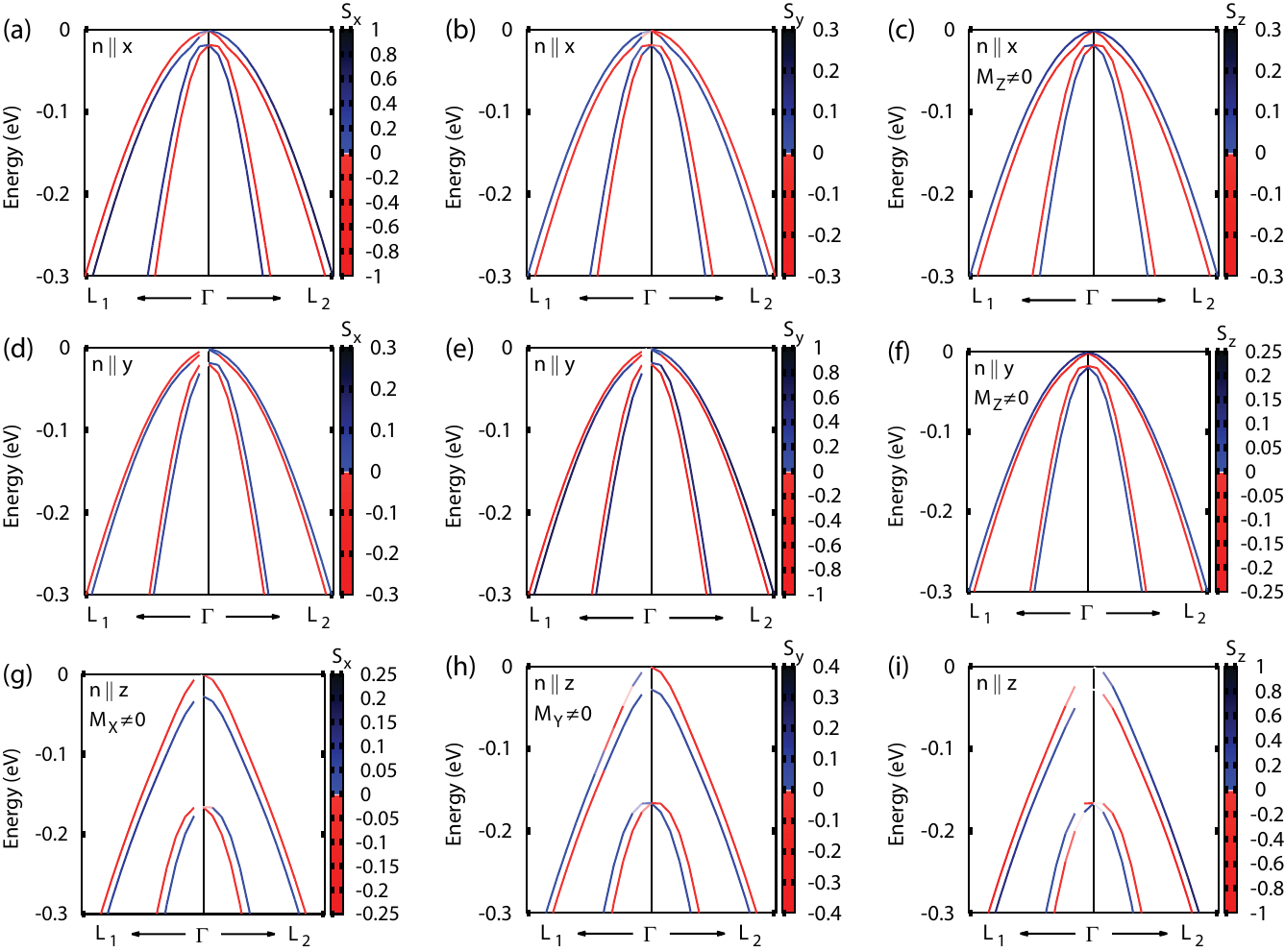}
\caption{Band structure of noncentrosymmetric MnSe with SOC along the L$_1$-$\Gamma$-L$_2$ low-symmetry k-path. With the N\'eel vector along the x-axis, we report the projection of the spin-component (a) S$_x$, (b) S$_y$ and (c) S$_z$. With the N\'eel vector along the y-axis, we report the projection of the spin-component (d) S$_x$, (e) S$_y$ and (f) S$_z$. With the N\'eel vector along the z-axis, we report the projection of the spin-component (g) S$_x$, (h) S$_y$ and (i) S$_z$. S$_x$, S$_y$ and S$_z$ represent the spin components for the total system. The Fermi level is set to zero. The coordinates of the k-points are L$_1$=(0.5,0.0.5) and L$_2$=(-0.5,0,0.5) in direct coordinates. When the \textbf{n} is along x, the total magnetization is M=(0,0,M$_z$). When the \textbf{n} is along y, the total magnetization is M=(0,0,M$_z$). When the \textbf{n} is along z, the total magnetization is M=(M$_X$,M$_Y$,M$_z$) where M$_z$ rises due to the weak ferrimagnetism.}
\label{L_MnSe}%
\end{figure*}

\begin{figure}[ht!]
\centering
\includegraphics[width=\linewidth]{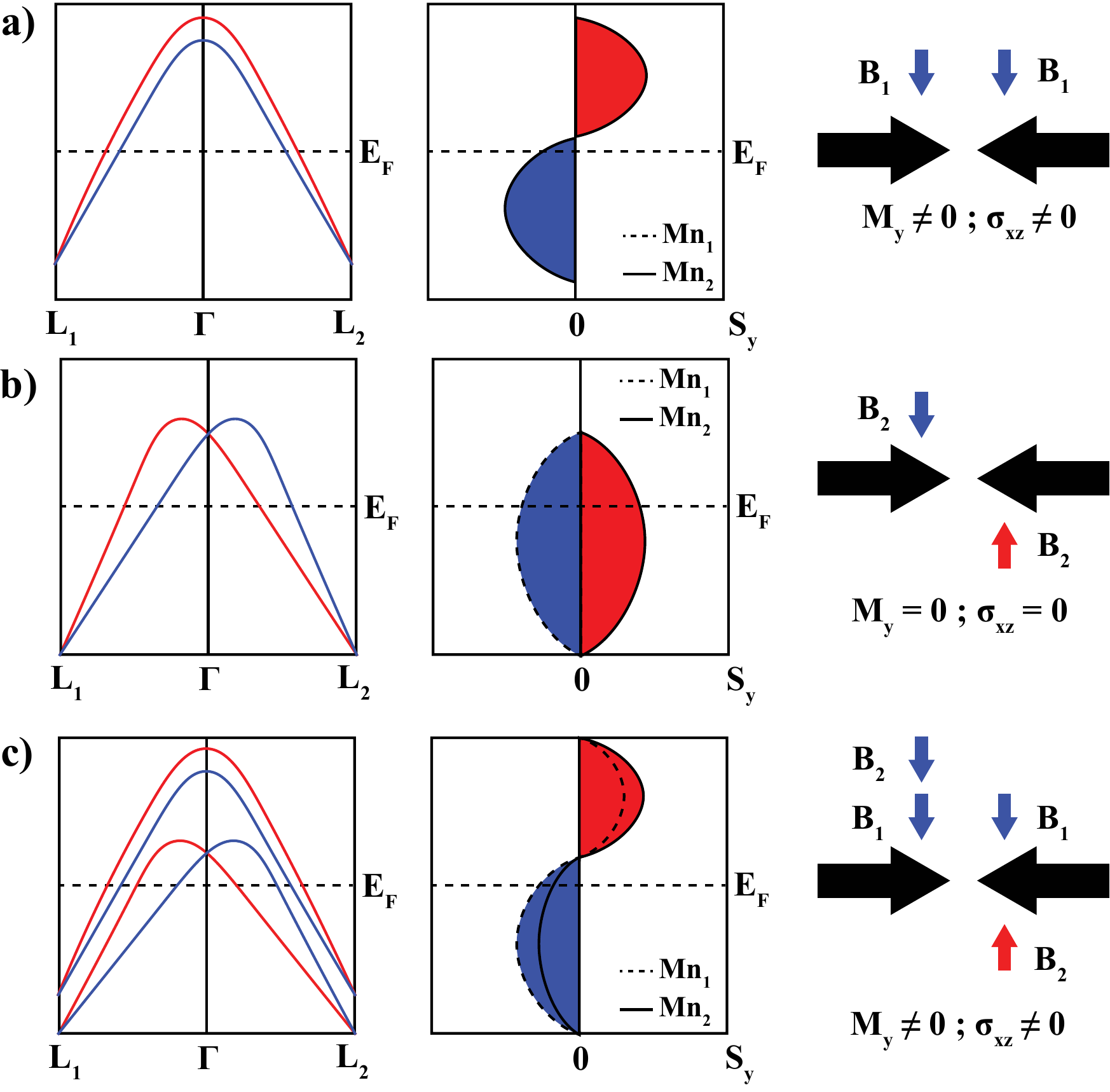}
\caption{Schematic band structure of noncentrosymmetric altermagnets with N\'eel vector along the z-axis. The red and blue colors represent the negative and positive S$_y$ components of the total system in the band structure. DOS of the spin component S$_y$ and magnetic configuration for three selected cases. a) Weak ferromagnetism in some bands (B$_1$) of noncentrosymmetric altermagnets. b) Rotation of the N\'eel vector in some bands (B$_2$) of noncentrosymmetric altermagnets. c) Weak ferrimagnetism as the superposition of weak FM and weak AFM from bands B$_1$ and B$_2$. }\label{weakferrimagnetism}
\end{figure}

\begin{figure}[ht!]
\centering
\includegraphics[width=\linewidth]{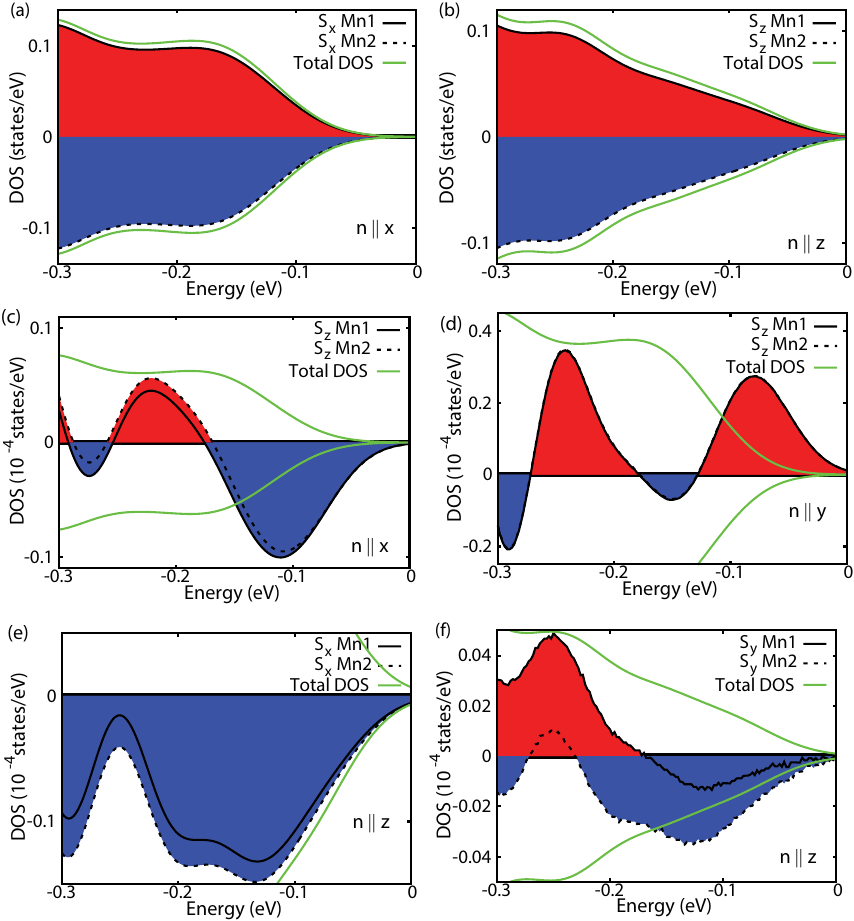}
\caption{DOS for the spin-components S$_x$, S$_y$ and  S$_z$ as a function of the energy for the wurtzite MnSe. (a,b) DOS for the spin components S$_x$ and S$_z$ with the N\'eel vector along the x-axis and along the z-axis, respectively. (c,d) DOS for the spin component S$_z$ with the N\'eel vector along the x-axis and along the y-axis, respectively. (e,f) DOS for the spin component S$_x$ and S$_y$
with the N\'eel vector along the z-axis, respectively. The solid and dashed lines represent the local spin components S$_x$, S$_y$ and S$_z$ for the two sublattices represented by Mn1 and Mn2.
The Fermi level is set to zero. The total DOS is renormalized by a factor of $10^{-4}$ states/eV} to fit within the plot. The negative values of the total DOS correspond to spin down while the positive values correspond to spin up.\label{DOS_MnSe}
\end{figure}

\begin{figure}[ht!]
\centering
\includegraphics[width=\linewidth]{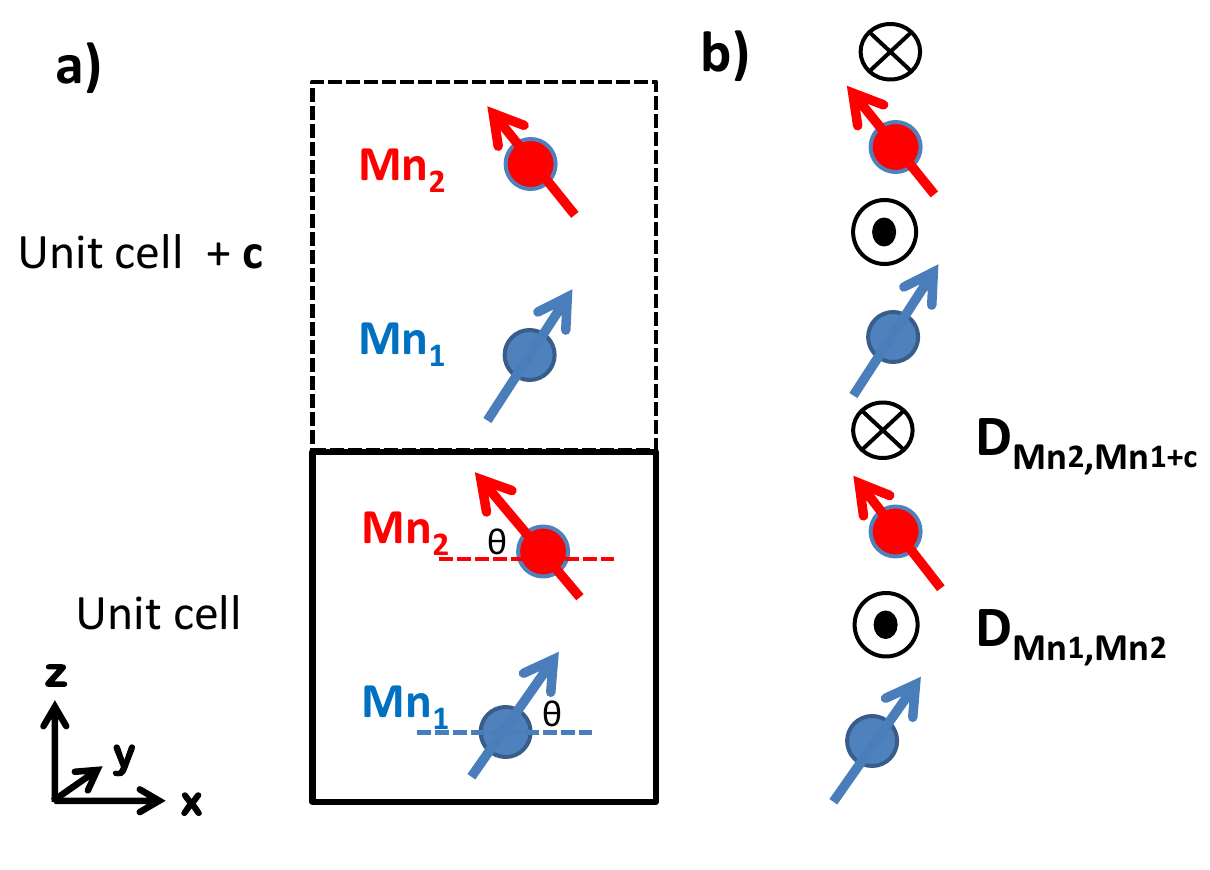}
\caption{Staggered DMI in the unit cell of MnTe with N\'eel vector along $\textbf{x}$. The blue (red) vectors show the spin up (spin down) sublattices Mn1 and Mn2 of the Mn atoms which are tilted around the x-axis by an angle $\theta$. (a) The spins in the primitive unit cell of MnTe and the repeated unit cell along the vertical c-direction and (b) show the cross-product vectors due to each pair of spins forming a staggered DMI along the c-direction.}\label{Staggered}
\end{figure}

\subsection{Weak FM and Weak ferrimagnetism in noncentrosymmetric hexagonal altermagnet MnSe}

To understand the effect of the breaking of the inversion symmetry, we analyze MnSe which is a hexagonal g-wave altermagnet like CrSb. Indeed, MnSe with a wurtzite crystal structure represents the case of an altermagnet without inversion symmetry with a small number of atoms in the unit cell. The structural and electronic properties without SOC for wurtzite MnSe are reported in the literature\cite{grzybowski2024wurtzite}. The non-relativistic spin-splitting in wurzite MnSe can reach a maximum of 200-300 meV in the Mn-bands where the SOC of Mn is of the order of 30 meV while the SOC of Se is 140 meV. Therefore, the SOC would be a perturbation at least in the region where the spin-splitting is maximum. For MnSe, the top of the valence band at $\Gamma$ and A are mainly p$_x$/p$_y$ with a sizeable contribution from Mn orbitals. When the N\'eel vector is moved from in-plane to out-of-plane, the top of the valence band rises. We detect a reduction of the band gap of around 80 meV (from 1.72 to 1.64 eV) when we move from AFM with the N\'eel vector in the plane to AFM with the N\'eel vector out-of-plane. The same sensitivity of the band gap from the N\'eel vector was found for the centrosymmetric g-wave altermagnet  MnTe.\cite{PhysRevB.107.L100417}

We calculate the band structure for the three different spin components along the low symmetry k-path L$_1$-$\Gamma$-L$_2$ where we have the non-relativistic spin-splitting. 
We find weak ferromagnetism for all N\'eel vectors. 
As in the case of CrSb, when the N\'eel vector is in the plane, the eigenvalues at the $\Gamma$ point split into two eigenvalues twofold degenerate with zero magnetization in the plane (see Fig. \ref{L_MnSe}(a,b,d,e)) and non-zero magnetization along the z-axis (see Figs.\ref{L_MnSe}(c,f)).\\
When the N\'eel vector is along the z-axis there are differences with respect to CrSb, there is a strong splitting with the two highest eigenvalues single-degenerate and the one eigenvalue double-degenerate at the $\Gamma$ point. The single-degenerate highest eigenvalues produce non-zero M$_x$ and M$_y$ as shown in Figs. \ref{L_MnSe}(g) and \ref{L_MnSe}(h), respectively. However, the double degenerate bands show antiparallel spin components.

Without SOC, there is a fourfold degeneracy in the top of the valence band\cite{grzybowski2024wurtzite}. When the system is noncentrosymmetric as in the wurzite structure, a combination of parallel and antiparallel spin orientation is possible despite both sets of bands coming from the same fourfold degenerate point without SOC. As a consequence in the DOS, there won't be a completely parallel or antiparallel DOS weight.
Finally, Fig. \ref{L_MnSe} (i) shows the sign and weight of S$_z$ on the two sublattices that are equal and opposite when the N\'eel vector is along the z-axis.
Qualitatively similar results are observed for the K$_1$-$\Gamma$-K$_2$ path (not reported) with the difference that the non-relativistic spin splitting is absent. 
We find weak ferromagnetism for g-wave compounds along the same direction with or without inversion symmetry. The main difference is the coexistence of both parallel and antiparallel magnetic configurations for the different bands of the in-plane spin-components when the N\'eel is along the z-axis. Basically, we have the superposition of weak ferromagnetism [see Fig. \ref{weakferrimagnetism}](a)] and rotation of the N\'eel vector [see Fig. \ref{weakferrimagnetism}](b)]. The superposition of both kinds of magnetism as reported schematically in Fig. \ref{weakferrimagnetism}(c) ends up forming weak ferrimagnetism (weak FiM). In this case, there is no exact compensation of the magnetization along the N\'eel vector and the Hall vector has a component along the N\'eel vector too.

While the altermagnets do not host non-relativistic spin-splitting at the $\Gamma$ point, we have observed a relativistic spin-splitting at the $\Gamma$ point in RuO$_2$ with N\'eel vector along the a-axis and in MnSe with N\'eel vector along the c-axis. In the case of weak ferromagnetism or weak ferrimagnetism, the spin-splitting at the time-reversal invariant points is possible. The spin-splitting at the time-reversal invariant points is necessary for a system to host the quantum anomalous Hall effect. Therefore, a possible route to achieve the quantum anomalous Hall effect involves having weak ferromagnets or weak ferrimagnets hosting relativistic spin-splitting at the $\Gamma$ point.

The DOSs for the spin components along the N\'eel vector are reported in Fig. \ref{DOS_MnSe}(a,b). The two sublattices, represented by the Mn$_1$ and Mn$_2$ local DOS, have equal and opposite DOS of the spin components since the spins are antiparallel.
The size of the DOS for the spin components orthogonal to the N\'eel vector is 3-4 orders of magnitude smaller than the size of the DOSs for the spin components along the N\'eel vector. When the N\'eel vector is along the x-axis, we observe from Fig. \ref{DOS_MnSe}(a,c) that the in-plane S$_x$ components of the two sublattices are antiparallel, while the spin components S$_z$ are parallel. Since the filling was completed and MnSe is an insulator, the weak ferromagnetism was not obtained in the calculations. However, assuming a p-doping, the magnetization M$_z$ obtained by integrating the spin density up to the Fermi level will be non-zero and the system will host a weak ferromagnetism along the z-direction. 
The same happens for the N\'eel vector along the y-axis, the in-plane components are antiparallel while we have weak ferromagnetism along S$_z$ as shown in Fig. \ref{DOS_MnSe}(d).
When the N\'eel vector is along the z-axis, we have both M$_x$ and M$_y$ components different from zero, as shown in Fig. \ref{DOS_MnSe}(e,f) with the Hall vector along the direction of the weak magnetization. We observe that the Hall vector is always orthogonal to the N\'eel vector. Even if some band structures resemble the Rahsba-like dispersions, the spin textures of the altermagnets are different and more complicated than the Rashba effect\cite{GUO2023100991}.
Of course, the camelback and a Rashba spin texture\cite{D3NJ01788E,Tao2021} rise on top of the altermagnetic electronic properties when the system lacks inversion symmetry like in the wurtzite structure.

\section{Staggered DMI interactions in altermagnets}

We show that in order to have weak ferromagnetism in altermagnets, it is sufficient to have a staggered DMI interaction. As a prototype, we choose the case of NiAs structure, the staggered DMI derives directly from the same rotations that connect the spin-up and spin-down sublattices in altermagnets\cite{Smejkal22}. We provide a concrete example of the consequence of the staggered DMI for the MnTe, CrSb and MnSe but the staggered DMI is present in several other altermagnetic compounds. 
Defining the x-axis as parallel to the first neighbor Mn-Mn distance and the y-axis as orthogonal to it in the ab plane (same notation used for CrSb in the previous Section), it was proved to have an anomalous Hall effect when the N\'eel vector is along the y-axis which is the easy axis for MnTe. Based on the presence of a mirror plane orthogonal to the y-axis, the DMI vector should have the y-component D$_y$ equal to zero. As a consequence, MnTe and CrSb should not be able to show AHE when the N\'eel vector is along the x-axis\cite{PhysRevLett.130.036702}.
We will show that the results reported in the previous Section for CrSb with different N\'eel vectors can be explained only with two DMI components that are D$_x$ and D$_y$. Therefore, in our results, we have a spontaneous breaking of the mirror plane. A recent paper supports this symmetry breaking due to the in-plane ferromagnetic order in the hexagonal structure, the interaction proposed to rise in MnTe is of higher order with respect to the DMI but it can approximated with a similar structure to the DMI\cite{mazin2024origingossamerferromagnetismmnte}. 
Since this higher-order interaction does not depend on the altermagnetism, this effect can appear in ferromagnetic phases with the same crystal structure.

\subsection{Staggered DMI and weak ferromagnetism in MnTe}

Experimentally, the N\'eel vector in MnTe is oriented along the y-axis, but we will analyze both cases with  N\'eel vector along the x-axis and y-axis. With the two spins in the unit cell on the two Mn atoms with the N\'eel vector along the x-axis or y-axis, the weak ferromagnetism rises along the z-axis as a consequence of the DMI. 
We define $\theta$ to be the canting angle measuring the deviation of the spins from the N\'eel vector once the DMI is introduced.
When the N\'eel vector is along $\mathbf{x}$, the canting angle $\theta$ is defined as in Figure \ref{Staggered}, therefore, the spins of Mn1 and Mn2 in the primitive unit cell are defined as $\vec{S}_{Mn1}$=(Scos($\theta$),0,Ssin($\theta$)) and $\vec{S}_{Mn2}$=(-Scos($\theta$),0,Ssin($\theta$)), respectively. If we shift the unit cell by the primitive vector \textbf{c}, we have the spin of the Mn1 atom with the same size and therefore $\vec{S}_{Mn1+\textbf{c}}$=(Scos($\theta$),0,Ssin($\theta$)). 
When the N\'eel vector is along $\mathbf{y}$, we need to exchange the x and y-coordinates in the spin vectors described above. 
Due to the mirror orthogonal to the c-axis which is not broken, the DMI vector lies in the xy plane, therefore, we have $\vec{D}_{Mn1,Mn2}$=(D$_x$,D$_y$,0) while the vector $\vec{D}_{Mn2,Mn1+\textbf{c}}$=(-D$_x$,-D$_y$,0) is opposite due to the rotation of 180 degree of the crystal structure. The same rotation of 180 degrees that gives rise to the non-relativistic spin-splitting also produces a staggered DMI. 
The spin vectors and the DMI vector are reported in Fig. \ref{Staggered}.
\\
The total magnetic energy of the system is
\begin{equation}
\begin{aligned}
    E = &J\vec{S}_{Mn1}\cdot\vec{S}_{Mn2} + J\vec{S}_{Mn2}\cdot\vec{S}_{Mn1+\textbf{c}} \\
    &+ \vec{D}_{Mn1,Mn2}\cdot(\vec{S}_{Mn1}\times\vec{S}_{Mn2}) \\
    &+ \vec{D}_{Mn2,Mn1+\textbf{c}}\cdot(\vec{S}_{Mn2}\times \vec{S}_{Mn1+\textbf{c}})
\end{aligned}
\end{equation}

The first two terms are the Heisenberg terms with J$>$0 for antiferromagnets, while the other two terms are the DMI interaction. The same terms of the magnetic energy between neighbours in the ab plane will bring to terms that are independent of $\theta$ and therefore just additive constants.
The total magnetic energy is therefore: 
\begin{equation}   
E(\theta)=-2JS^2cos(2\theta)+2D_yS^2sin(2\theta)\quad if \quad \boldsymbol{n} ||  \boldsymbol{x}
\end{equation}
\begin{equation}   
E(\theta)=-2JS^2cos(2\theta)+2D_xS^2sin(2\theta)\quad if \quad \boldsymbol{n} ||  \boldsymbol{y}
\end{equation}
When the N\'eel vector is along x (y), the component D$_x$ (D$_y$) does not play any role but in general, it is present.  Therefore, the equilibrium canting angle $\overline{\theta}$ is the one minimizing the magnetic energy in equation (2). We obtain the pivotal role played by the staggered DMI interaction $D_y$ (focusing on the case $\boldsymbol{n} ||  \boldsymbol{x}$) present in altermagnets 
\begin{equation}\label{DJ}
2\overline{\theta}\approx\tan(2\overline{\theta})=-\frac{D_y}{J}
\end{equation}
where we can approximate the trigonometric functions for small angles since the magnetic exchange J is usually much larger than the DMI vector. We can separate the magnetization in the xy plane component M$_{plane}$ and the magnetization along the z-axis M$_z$. They are given by: 
\begin{equation}\label{MZ}
M_z=2S\sin(\overline{\theta}) \approx 2S\overline{\theta} \approx -\frac{SD_y}{J}; \quad M_{plane} =0
\end{equation}
From equation (\ref{MZ}), we obtain that the weak magnetization depends on D/J. $\overline{\theta}$ is the canting angle of the bulk weak ferromagnet. 
The equations describing the canting angle in altermagnets with staggered DMI are the same exact ones describing spin-spirals in ferromagnets with a uniform DMI. We can think of tuning the DMI in altermagnets with the same strategy proposed to tune the DMI in ferromagnetic layers but taking into account the staggered geometry. We can cover with 5d heavy metals, among these Re and Pt produce opposite chiralities in Co thin films, therefore, one would increase and another would decrease the DMI. \cite{10.1063/5.0177260}
Covering with Pt (or Re) increases the DMI, the Pt (or Re) covering the bottom or the top of a thin film will produce the opposite DMI. Differently from the chiral multilayer systems, we will have an odd-even effect with the odd and even number of layers behaving differently due to the staggered DMI.

\begin{figure*}[ht!]
\centering
\includegraphics[width=\linewidth]{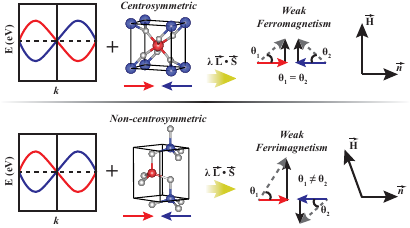}
\caption{Once altermagnetism breaks the time-reversal symmetry, the spin-orbit $\lambda\Vec{L}\cdot\Vec{S}$ can produce weak ferromagnetism (top panel) or weak ferrimagnetism (bottom panel) depending on the crystal symmetries and N\'eel vector orientation.
It seems that weak ferromagnetism tends to be present in high-symmetric cases as centrosymmetric crystals, while weak ferromagnetism tends to be present in low-symmetry cases as non-centrosymmetric crystals.
In the first case, the Hall vector $\Vec{H}$ is orthogonal to the N\'eel vector $\Vec{n}$ while in the second case, the Hall vector $\Vec{H}$ has a component parallel to the N\'eel vector $\Vec{n}$. $\theta_1$ and $\theta_2$ are the canting angles for the two magnetic atoms. Although other magnetic configurations are possible for weak ferromagnetism and weak ferrimagnetism, the two above presented could be among the most common.}\label{grap}
\end{figure*}

\subsection{Staggered DMI and weak ferromagnetism in CrSb}

CrSb is also in the NiAs structure but the experimental N\'eel vector is along $\mathbf{z}$.
The total magnetic energy for CrSb is analogous to MnTe:
\begin{equation}
\begin{aligned}
    E = &J\vec{S}_{Cr1}\cdot\vec{S}_{Cr2} + J\vec{S}_{Cr2}\cdot\vec{S}_{Cr1+\textbf{c}} \\
    &+ \vec{D}_{Cr1,Cr2}\cdot(\vec{S}_{Cr1}\times\vec{S}_{Cr2}) \\
    &+ \vec{D}_{Cr2,Cr1+\textbf{c}}\cdot(\vec{S}_{Cr2}\times \vec{S}_{Cr1+\textbf{c}})
\end{aligned}
\end{equation}
When the N\'eel vector is along $\mathbf{z}$, we take $\theta$ to be the canting angle while $\phi$ is the azimuthal angle, and therefore, the spins of Cr1 and Cr2 in the primitive unit cell are defined as $\vec{S}_{Cr1}$=(Scos($\phi$)sin($\theta$),Ssin($\phi$)sin($\theta$),Scos($\theta$)) and $\vec{S}_{Cr2}$=(Scos($\phi$)sin($\theta$),Ssin($\phi$)sin($\theta$),-Scos($\theta$)), respectively. The total magnetic energy if $\textbf{n} || \textbf{z}$ is therefore:
\begin{equation}   
E(\theta,\phi)=-2JS^2cos(2\theta)+2(D_xcos(\phi)+D_ysin(\phi))S^2sin(2\theta)
\end{equation}
with D$_x$ and D$_y$ that are not zero.

When we impose the derivative respect to phi equal to zero:
\begin{equation}   
\partial_{\phi}E(\theta,\phi)= 2(-D_x\sin(\phi)+D_y\cos(\phi))S^2sin(2\theta)=0
\end{equation}
we get the solutions $\theta$=0 and $\theta$=$\pi$/2 that are the collinear antiferromagnet and the collinear ferromagnet, respectively.
The solution for which we have the minimum of the energy is 
\begin{equation}\label{azimuthal}   
\tan(\phi)=\frac{D_y}{D_x}
\end{equation}
where the angle $\phi$ is not necessarily small.
When we impose the derivative respect to theta equal to zero
\begin{equation}   
\partial_{\theta}E(\theta,\phi)=0
\end{equation}
we get
\begin{equation}
\tan(2\overline{\theta})=-\frac{(D_x^2+D_y^2)\sin(\overline{\phi})}{JD_y}
\end{equation}
and using equation (\ref{azimuthal}) (assuming $\phi$ between 0 and $\pi$/2) we get:
\begin{equation}
\tan(2\overline{\theta})=-\frac{\sqrt{D_x^2+D_y^2}}{J}
\end{equation}
this equation is the analog to equation (\ref{DJ}) where we have at the numerator the module of DMI excluding the component parallel to the N\'eel vector and the magnetic coupling at the denominator that makes in most of the cases $\overline{\theta}$ very small. Therefore, we have the magnetization components to be:
\begin{equation}
M_{plane} \approx 2S\overline{\theta} \approx -\frac{S\sqrt{D_x^2+D_y^2}}{J}; \quad M_{z} = 0
\end{equation}

\subsection{Weak ferrimagnetism in MnSe}

Moving to the noncentrosymmetric compounds, the crystal structure has a lower symmetry and this would make the structure of DMI vectors more complicated and it produces two different canting angles for the two magnetic atoms. 
When the N\'eel vector is in the plane, the spins on the two Mn atoms are equivalent and we obtain for MnSe the same weak ferromagnetism of CrSb. When the N\'eel vector is out of the plane, the spins on the two Mn atoms are inequivalent. The schematic pictures of weak ferromagnetism and weak ferrimagnetism are reported in Fig. \ref{grap}, where a component of the weak ferrimagnetism rises parallel to the N\'eel vector due to different canting angles ${\theta}_1$$\neq{\theta}_2$ of the two magnetic atoms.
Therefore, the weak ferrimagnetism of the MnSe with N\'eel vector along the z-axis, the component parallel to the N\'eel vector will be M$_z$=S$\cos(\overline{\theta}_1)$-S$\cos(\overline{\theta}_2)$. In the limit of small $\overline{\theta}_1$ and $\overline{\theta}_2$, this becomes  M$_z\approx\frac{S}{2}$($\overline{\theta}_2^2$-$\overline{\theta}_1^2$). Assuming that $\overline{\theta}_1$ and $\overline{\theta}_2$ depend on the ratio D/J as well, we have that in the weak ferrimagnetism the component parallel to the N\'eel vector M$_z$ is proportional to the square of the DMI in the form M$_z\approx\frac{S}{2}$($\frac{D^2}{J^2}$). Considering that M$_{plane}\approx$  $-\frac{S\sqrt{D_x^2+D_y^2}}{J}$, both in-plane and out-of-plane magnetic components are non zero.

\section{Discussion and Conclusions}

In this paper, we propose that all compounds hosting weak ferromagnetism induced by the relativistic Dzyaloshinskii-Moriya interaction\cite{DZYALOSHINSKY1958241,PhysRev.120.91} belong to the family of the altermagnets. While Dzyaloshinskii and Moriya focused on the interplay between magnetic interactions and symmetries, recent studies on altermagnetism provided a deeper knowledge of the crystal symmetries, extended the investigation on the magnetic properties and discovered new effects on the electronic properties\cite{mazin2022altermagnetism,hayami2019momentum,hayami2020bottom,Smejkal22,Smejkal22beyond, Fakhredine23}. The connection between these two fields emphasized in this paper paves the way to simplify the physical scenario and merge the two fields giving the chance to connect the literature developed from two different perspectives which until now had a minor overlap.

\begin{table}[th]
\centering
\setlength{\tabcolsep}{7 pt}
\renewcommand{\arraystretch}{1}
\scalebox{1.1}{
\begin{tabular}{|c|c|c|}
\hline
Compound  & NO SOC & Weak FM \\
(No. space group)   & splitting & from DMI \\
\hline
(Sr,Ca)$_3$Ru$_2$O$_7$(36) & - & Exp.\cite{PhysRevB.57.978} \\ 
Ca$_3$Mn$_2$O$_7$ (36) & Th.\cite{GUO2023100991} & Exp.\cite{sg-36} \\ 
\hline
Mn(N(CN)$_2$)$_2$ (58) & Th.\cite{GUO2023100991} & Exp.\cite{sg-58}\\
\hline
YVO$_3$ (62) & Th.\cite{Cuono23orbital}  & Exp.\cite{PhysRevB.62.6577} \\
LaMnO$_3$ (62) & Th.\cite{D3NR03681B} & Exp.\cite{weakFM_LAO} \\
CaMnO$_3$ (62)& Th.\cite{GUO2023100991} & Exp.\cite{macchesney1967magnetic} \\
YCrO$_3$ (62)& Th.\cite{GUO2023100991} & Exp.\cite{JUDIN1966661} \\
CaCrO$_3$ (62)& Th.\cite{PhysRevB.107.155126} & Th.\cite{PhysRevB.107.155126} \\
\hline
Ba$_2$CoGe$_2$O$_7$ (113)&  Th. \cite{GUO2023100991} & Exp.\cite{sg-113} \\
\hline 
RuO$_2$ (136) & Th.\cite{Smejkal22} & Th.*, Exp.\cite{Feng2022} \\
\hline
La$_2$NiO$_4$ (138) & Th.\cite{GUO2023100991}  & Exp.\cite{sg-138} \\
\hline
LiFeP$_2$O$_7$ (140) & Th.\cite{GUO2023100991}  & Exp.\cite{sg-140-a} \\
KMnF$_3$ (140) & Th.\cite{GUO2023100991} & Exp.\cite{sg-140-b} \\
\hline
MnTiO$_3$ (161) & Th.\cite{GUO2023100991} & Exp.\cite{PhysRevB.88.104416} \\ 
\hline
$\alpha$-Fe$_2$O$_3$ (167) & Th.\cite{GUO2023100991}  & Th.\cite{PhysRev.120.91}, Exp.\cite{Fe2O3-expt} \\
MnCO$_3$ (167) & Th.\cite{GUO2023100991}  & Th.\cite{PhysRev.120.91}, Exp.\cite{MnCO3-expt} \\
CoCO$_3$ (167) & Th.\cite{GUO2023100991}  & Th.\cite{PhysRev.120.91}, Exp.\cite{CoCO3-expt} \\
CrF$_3$ (167) & Th.\cite{gao2023aiaccelerated}  & Th.\cite{PhysRev.120.91}, Exp.\cite{CrF3-expt-1} \\
CoF$_3$ (167) & Th.\cite{gao2023aiaccelerated}  & Exp.\cite{LEE201894} \\
FeF$_3$ (167) & Th.\cite{gao2023aiaccelerated}  & Exp.\cite{LEE201894} \\
FeBO$_3$ (167) & Th.\cite{GUO2023100991}  & Exp.\cite{FeBO3-expt2} \\
\hline
Ni$_{1/3}$NbS$_2$ (182) & Th.\cite{Sattigeri24}  &  - \\
Ni$_{1/3}$TaS$_2$ (182) & Th.\cite{Sattigeri24}  &  - \\
Co$_{1/3}$NbS$_2$ (182) &         -              & Exp.\cite{PhysRevB.105.L121102}\\
\hline 
MnSe (186) & Th.\cite{grzybowski2024wurtzite} & Th.* \\
\hline
MnTe (194) & Th.\cite{Smejkal22}, Exp.\cite{krempaský2023altermagnetic} & Th.\cite{kluczyk2023coexistence}, Exp.\cite{kluczyk2023coexistence} \\
CrSb (194) & Th.\cite{Smejkal22}, Exp.\cite{reimers2024direct} & Th.*, Exp.\cite{CrSb-expt} \\
\hline
\end{tabular}}
* Present work.\hspace{6cm}
\caption{List of selected altermagnetic compounds where the non-relativistic spin-splitting appears with weak ferromagnetism induced by DMI. The words "Th." and "Exp." are reported for the theoretical and experimental verification of the non-relativistic spin splitting or weak ferromagnetism. The space groups 36, 182 and 186 lack inversion symmetry.}
\label{Table1}
\end{table}

In Table \ref{Table1}, we report several altermagnetic compounds with different space groups where the non-relativistic spin-splitting and the weak ferromagnetism were theoretically predicted or experimentally detected. All the compounds described in the seminal papers of Dzyaloshinskii and Moryia are altermagnets present in this table\cite{PhysRev.120.91}. 
Due to the lifting of the Kramers degeneracy, all altermagnetic systems could potentially show weak ferromagnetism, however, for particular N\'eel vector orientations the DMI is zero as in the case of the RuO$_2$ with N\'eel vector along the z-axis.
The g-wave CrSb exhibits weak ferromagnetism for every N\'eel vector. On the other hand, there are altermagnets which display weak ferromagnetism only for a given N\'eel vector, such as the d-wave CaCrO$_3$\cite{PhysRevB.107.155126} and RuO$_2$. Indeed, for RuO$_2$ we have shown that there is no weak FM if the N\'eel vector is oriented along the z-axis. Finally, if we consider the compound MnTe$_2$\cite{zhu2024observation,Autieri2024} or the A-centered magnetic phase of Ca$_2$RuO$_4$ with N\'eel vector along the b-axis\cite{Cuono25}, we have no weak ferromagnetism since the cantings compensate producing zero net magnetization along all directions.
As a consequence of this observation, the family of altermagnets is wider than the family of weak ferromagnets since there are altermagnets that do not exhibit weak ferromagnetism.
For the noncentrosymmetric systems, we have found that weak FM or weak FiM are present depending on the N\'eel vector orientation.

Beyond the crystal symmetries and the N\'eel vector orientation, the only other requirement for weak FM or weak FiM is to have a non-compensated magnetization as happens in metallic systems or doped semiconductors. As shown in Fig. \ref{Fig1}, when the system is insulating no weak ferromagnetism is observable since there is a spin compensation, therefore, a small doping is necessary like in the case of MnTe\cite{kluczyk2023coexistence} and other semiconductive altermagnets.
From the DOS of the spin components, we can understand that doping can change the Fermi level, strongly modify the weak ferromagnetism, and can even reverse the sign of the DMI. This explains the dependence of the weak ferromagnetism from electron- or hole-doping concentration in several cases\cite{KHOMCHENKO20081927,PhysRevLett.119.167201}. What was defined as "weak altermagnetism" is the band structure effect that gives rise to the DMI in centrosymmetric altermagnets.\cite{krempaský2023altermagnetic} 
The electric field increases or activates the non-relativistic spin-splitting depending on the field orientation\cite{D3NR03681B,mazin2023induced}, therefore, the electric field can increase and manipulate the DMI in altermagnets\cite{PhysRevLett.100.167203}. 
We can make another observation regarding centrosymmetric altermagnetic systems that can host an active DMI. In the case of magnetic systems with frustrated magnetic interactions, the presence of an active DMI in altermagnetic crystal structure can further push the system towards helical magnetic structures, spin spiral and complex magnetism as observed for instance in transition metal monopicnitides\cite{Wang2016,Campbell2021}. 
The component of the weak ferromagnetism is proportional to the spin-orbit, therefore, to look for a large canting angle we need to search among systems with a strong spin-orbit and a small magnetic moment.
In weak ferromagnets, an extremely relevant role is given by the magnetocrystalline anisotropy (MCA) which decides a key ingredient as the direction of the N\'eel vector. Additionally, the size of the MCA determines how easy it is for the spin to rotate, therefore a low MCA is fundamental to favor the weak ferromagnetism. Even if there are altermagnets where the weak ferromagnetism is zero in the bulk as the compounds belonging to the second class, the weak ferromagnetism can be obtained by breaking crystal symmetries on the surfaces. Further studies are necessary for the evolution of the DMI and weak ferromagnetism on the surfaces of the altermagnets.\cite{pylypovskyi2023surfacesymmetrydriven,D3NR03681B}


In conclusion, we propose that weak ferromagnetism induced by staggered DMI can appear in centrosymmetric and noncentrosymmetric altermagnetic compounds but not in centrosymmetric and noncentrosymmetric conventional antiferromagnets. 
The weak ferromagnetism and the AHE are orthogonal to the N\'eel vector for centrosymmetric systems for the cases considered here, even if it is not always the case. The weak ferromagnetism depends on the filling, could be tuned by an electric field and can change its sign by rotating the N\'eel vector in hexagonal systems. From our results, the weak ferromagnetism is absent for MnSe seminconductor with filled shells, but it could be activated with light doping and then the weak ferromagnetism would be proportional to the carriers.
We provide a recipe to establish the presence of weak ferromagnetism and the components of the Hall vector from a simple relativistic density of state and band structure calculations without calculating the DMI vector. There is a correlation between the rise of weak FM and weak FiM with the lifting of the electronic degeneracy at the $\Gamma$ point. The centrosymmetric altermagnets can host different cases: the altermagnets can host weak ferromagnetism for every direction of the Hall vector while in other cases the altermagnets can host weak ferromagnetism for suitable directions of the N\'eel vector. In a third case, the altermagnets do not host weak ferromagnetism. For noncentrosymmetric MnSe systems, we have weak ferromagnetism in the x-y plane while parallel and antiparallel spin components for different bands can coexist at the same energy resulting in weak ferrimagnetism and a component of the Hall vector along the N\'eel vector. We have predicted weak ferromagnetism in CrSb and weak ferrimagnetism in wurztite MnSe. The size of weak ferromagnetism, weak ferrimagnetism and AHE strongly depends on the spin-orbit coupling, therefore, further research directions should focus on heavy elements. In summary, the family of the altermagnets is mainly composed of weak ferromagnets with a small number of compounds which are weak ferrimagnets (MnSe with N\'eel vector along the z-axis) and zero magnetization compounds (RuO$_2$ with N\'eel vector along the z-axis).\\
The breaking of the time-reversal symmetry (altermagnetic spin-splitting) is a necessary but not sufficient condition to obtain the relativistic weak ferromagnetism or weak ferrimagnetism. 
The weak FiM tends to appear in the lower symmetry cases, therefore, an external electric field breaking the inversion symmetry could generate a weak ferrimagnetism in suitable cases. The rise of weak ferromagnetism or weak ferrimagnetism depends on the interplay between the space group, magnetic order, and the N\'eel vector orientation. In the case of MnTe, the weak ferromagnetism is produced by the staggered DMI where the property of the staggered DMI comes down from the very same rotations responsible for the altermagnetism in this compound.
While the weak ferromagnetism has net magnetization with components linearly proportional to the D/J, the weak ferrimagnetism in MnSe also shows components proportional to the second order in D/J.\\

\begin{acknowledgments}
We acknowledge M. Cuoco, K. V\'yborn\'y,  M. 
Gryglas-Borysiewicz and M. Grzybowski for useful discussions. This research was supported by the Foundation for Polish Science project "MagTop" no. FENG.02.01-IP.05-0028/23 co-financed by the European Union from the funds of Priority 2 of the European Funds for a Smart Economy Program 2021–2027 (FENG). A.F. was supported by the Polish National Science Centre under project no. 2020/37/B/ST5/02299. We acknowledge the access to the computing facilities of the Interdisciplinary Center of Modeling at the University of Warsaw, Grant g91-1418, g91-1419, g91-1426, g96-1808 and g96-1809 for the availability of high-performance computing resources and support. We acknowledge the CINECA award under the ISCRA initiative IsC99 "SILENTS”, IsC105 "SILENTSG", IsB26 "SHINY" and IsB27 "SLAM" grants for the availability of high-performance computing resources and support. We acknowledge the access to the computing facilities of the Poznan Supercomputing and Networking Center Grant No. 609, pl0223-01 and  pl0267-01.
\end{acknowledgments}

\appendix

\begin{figure}[ht!]
\includegraphics[width=\linewidth]{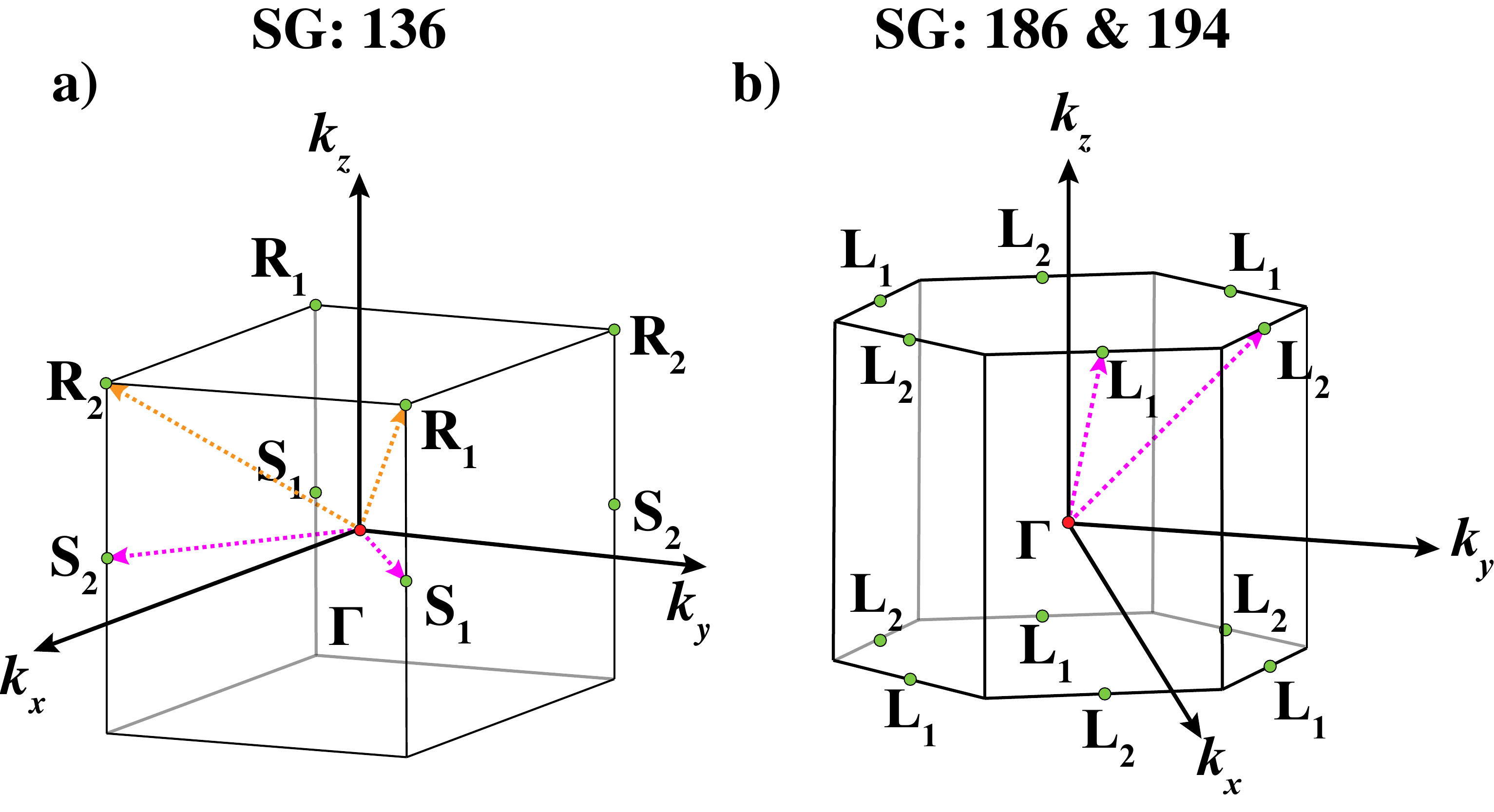}
\caption{(a) Brillouin zone of RuO$_2$ belonging to space group numbers 136.
(b) Brillouin zone of MnSe and CrSb belonging to the space group 186 and 194, respectively. The high-symmetry k-points used in this paper have been indicated.}\label{brillouin-zone}
\end{figure}

\section{Computational details}
The electronic properties were computed within the framework of density functional theory calculations based on plane wave basis set and projector augmented wave method using VASP\cite{Kresse93,Kresse96,Kresse96b} package. A plane-wave energy cut-off of 400~eV was used and the generalized gradient approximation of Perdrew, Burke, and Ernzerhof has been used\cite{perdew1996generalized}. The Hubbard U effects for the 4d-orbitals of Ru$^{4+}$, the 3d-orbitals of Cr and 3d-orbitals of Mn$^{2+}$ have been adopted\cite{Liechtenstein95density}. We have used the value of U = 2 eV \cite{Ivanov2016,PhysRevResearch.4.023256} keeping J$_H$ = 0.15U. The value of U strengthens the gap in MnSe and establishes a robust magnetic moment in RuO$_2$, however, all qualitative results are independent of the value of U.
We have performed the calculations using 20$\times$20$\times$28 and 24$\times$24$\times$18 k-points centered at $\Gamma$ for RuO$_2$ and CrSb, respectively. A denser k-grid of 28$\times$28$\times$18 was used for wurtzite MnSe, which is necessary to have an accurate description of the DOS, especially in the presence of relatively small spin-orbit coupling (SOC). Minor numerical errors of the order of 10$^{-5}$ states/eV are present in the DOS and in the magnetization of the band structure since we must switch off the symmetries during the relativistic calculations.\\

The lattice constants are a=4.4825 {\AA}  and c=3.1113 {\AA} for the RuO$_2$,
a=4.103 {\AA} and c=5.463 {\AA}\cite{reimers2024direct}  for the CrSb and a=4.12 {\AA} with c=6.72 {\AA} for the MnSe\cite{grzybowski2024wurtzite}. The experimental N\'eel vectors are along the z-axis for RuO$_2$ and CrSb\cite{D0DT03277H}, while it is experimentally unknown for MnSe but ab initio results predict that the N\'eel vector of MnSe is in the ab plane.\cite{grzybowski2024wurtzite}
All calculations were performed in the presence of SOC. The information on the individual spinor components is available for the site's projected density of states in the VASP code. The site-projected orbital- and energy-resolved spin density S$_x$, S$_y$ and S$_z$ are available. We sum over all orbitals to obtain the total S$_x$, S$_y$ and S$_z$ components. The smearing value for the DOS was 50 meV. 

Regarding the magnetic space group (MSG), the RuO$_2$ has MSGs 58.398 and 136.499 for the N\'eel vector along the x-axis and z-axis, respectively. 
CrSb has MSGs 63.457, 63.462 and 194.268 for the N\'eel vector along the x-axis, y-axis, and z-axis, respectively. 
The magnetic point group without spin-orbit coupling\cite{turek2022altermagnetism,PhysRevB.102.014422} of the wurtzite MnSe is 6$'$m$'$m, different from that of the hexagonal CrSb in the NiAs structure which is 6$'$/m$'$m$'$m. However, both groups belong to the same magnetic Laue class (6$'$/m$'$m$'$m) proving that the spin splitting of the electronic bands without SOC is qualitatively the same in both systems which have the same Brllouine zone. This could be established using the formalism of spin groups\cite{Smejkal22,Smejkal22beyond}. 
The Brillouin zone for RuO$_2$ is reported in Fig. \ref{brillouin-zone}(a), while the Brillouin zone for CrSb and MnSe is the same hexagonal Brillouin zone reported in Fig. \ref{brillouin-zone}(b).

\section{Materials candidate for large canting angles}

It seems that there are very few cases where the canting angles can be large, among these cases we can mention the iridates oxides.
These same altermagnetic properties valid for RuO$_2$ are also valid for the large spin-orbit IrO$_2$\cite{ming2017metalinsulator}, the main difference is that IrO$_2$ becomes magnetic just in thin films due to a transition driven by the bandwidth reduction. Analogously, also SrIrO$_3$ (space group 62) in the ultrathin limit and Sr$_2$IrO$_4$ 
($I41/a$, space group no. 88) can show weak ferromagnetism in centrosymmetric systems with canting angles up to 12.2$^{\circ}$\cite{Boseggia2013}.

\section{Magnetic energy express as a function of the N\'eel vector}

Instead of the formalism used in equations (2) or (3), another formalism from the literature\cite{PhysRevB.71.060401} was used for the altermagnet BiFeO$_3$. 
Defining \textbf{n} and \textbf{M} as the N\'eel vector and the sum of the spin we can write the total energy as:
\begin{equation}
E_{DM}= \vec{D}\cdot(\vec{n}{\times}\vec{M})
\end{equation}
where we can observe how the magnetic energy strongly depends on the N\'eel vector orientation.

\bibliography{altermagnetism}
\end{document}